\documentclass[english,twocolumn,tightenlines,showpacs,preprintnumbers,amsmath,amssymb,prl]{revtex4}

\usepackage{amsmath}
\usepackage{graphicx}
\usepackage{amssymb}

\begin{document}


\title{Efficient routing of single photons by one atom and a microtoroidal cavity}
\author{Takao Aoki,$^{\dagger}$$^{a}$ A. S. Parkins,$^{b}$ D. J. Alton,$^{\dagger}$ C. A. Regal,$^{\dagger}$ Barak Dayan,$^{\dagger}$$^{c}$ E. Ostby,$^{\ddag}$ K. J. Vahala,$^{\ddag}$ and H. J. Kimble$^{\dagger}$}
\affiliation{$^{\dagger}$Norman Bridge Laboratory of Physics 12-33, $^{\ddag}$T. J. Watson Laboratory of Applied Physics, California Institute of
Technology, Pasadena, California 91125, USA}

\date{\today}

\begin{abstract}
Single photons from a coherent input are efficiently redirected to
a separate output by way of a fiber-coupled microtoroidal cavity
interacting with individual Cesium atoms. By operating in an overcoupled
regime for the input-output to a tapered fiber, our system
functions as a quantum router with high efficiency for photon sorting. 
Single photons are reflected and excess photons transmitted, 
as confirmed by observations of photon antibunching (bunching)
for the reflected (transmitted) light. Our photon router is robust
against large variations of atomic position and input power, with
the observed photon antibunching persisting for intracavity photon
number $0.03\lesssim\bar{n}\lesssim0.7$. 
\end{abstract}


\maketitle

Cavity quantum electrodynamics (cQED) offers systems in which the
coherent interaction between matter and light can dominate irreversible
channels of dissipation~\cite{Vahala04,Khitrova06,Schoelkopf08,Kimble08}. 
Diverse systems based upon radiative interactions in cQED are thereby promising
candidates for the physical implementation of quantum networks, where,
for example, atoms in optical cavities (quantum nodes) 
are linked by photons in optical fiber (quantum channels)~\cite{Kimble08}.
Although many important capabilities for quantum nodes have been demonstrated
within the setting of cQED with single atoms in Fabry-Perot 
cavities~\cite{Turchette95,McKeever04,Keller04,Hijelkema07,Boozer07,Wilk07}, 
an outstanding challenge is high efficiency transport of quantum fields into and out of optical cavities~\cite{Kimble08},
as is required to link large numbers of quantum nodes.

In this regard, the coupling rate $\kappa$ of photons to and from
the quantum channel should dominate the rates for any other input-output
mechanisms. One way to achieve this is to operate the nodes
in an overcoupled regime~\cite{Spillane03}, where
external coupling dominates internal system losses. In the microwave
domain, ``circuit QED'' systems routinely operate in the overcoupled
regime~\cite{Schoelkopf08} for high external efficiency.

In this Letter, we realize a cQED system in the optical domain with
efficient input-output coupling while still maintaining high internal
efficiency for coupling to a single atom. We use a microtoroidal cavity
interacting with single Cesium atoms~\cite{Armani03,Aoki06,Dayan08}, with
coupling to and from the cavity implemented with
a tapered optical fiber in an overcoupled regime~\cite{Spillane03}. 
As a proof of principle, we demonstrate an efficient and robust
photon router for which single photons are extracted from an incident
coherent state and redirected to a separate output with efficiency
$\xi\simeq0.6$.

\begin{figure}[t]
 \includegraphics[width=8.6cm]{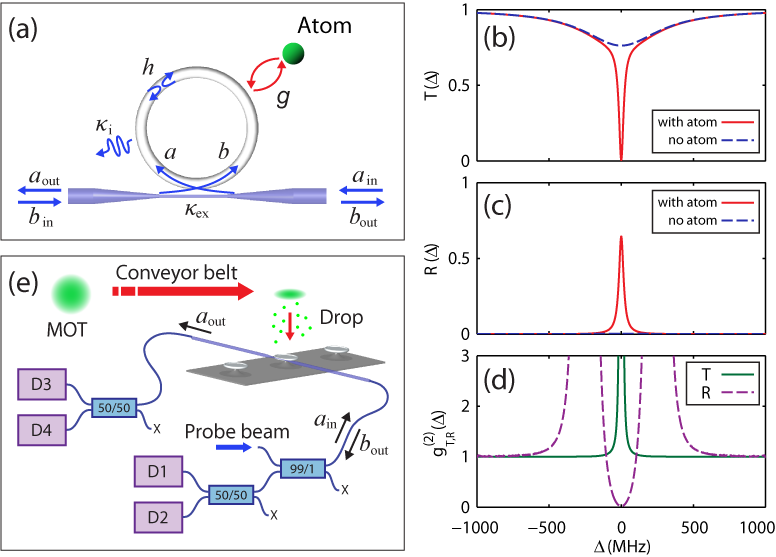}
\caption{\label{fig:1} (a) Simple depiction of one atom coupled to a toroidal
cavity, together with fiber taper and relevant field modes, with
rates $(g_{\rm tw},\kappa_{{\rm ex}},\kappa_{{\rm i}},h)$ as defined in the
text. (b-d) Theoretical plots for the parameters of our experiment,
$(g_{\rm tw},\kappa_{{\rm ex}},\kappa_{{\rm i}},h)/2\pi=(50,300,20,10)$ MHz,
with $\omega_{{\rm A}}=\omega_{{\rm C}}$. (b, c) Transmission and
reflection spectra $T(\Delta),R(\Delta)$ for $a_{out},b_{out}$ as
functions of probe detuning $\Delta=\omega_{C}-\omega_{{\rm p}}$
with and without the atom. (d) Theoretical intensity correlation functions versus
$\Delta$ for the transmitted ($g_{T}^{(2)}(\tau=0)$) and reflected
($g_{R}^{(2)}(\tau=0)$) fields. (e) Schematic of our experiment.
A Cesium MOT is formed in a separate chamber, and atoms are transfered
to the main chamber via an optical conveyor belt. Atoms are then dropped
onto a toroid, which is coupled to a tapered fiber as in (a).
A probe beam $a_{in}$ is injected into the taper, and the transmitted
$a_{out}$ and reflected $b_{out}$ fields are measured with two pairs
of single-photon detectors $D_{1-4}$. }
\end{figure}

To model photon transport for the atom-cavity system, we consider
the interaction of one atom with the evanescent fields of a microtoroidal
cavity, as shown in Fig.~1(a), with $g_{\rm tw}$ the rate of coherent
atom-cavity coupling \cite{g}. Near the atomic resonance at frequency $\omega_{A}$, the cavity
supports two counter-propagating modes $a,b$ of frequency $\omega_{C}$.
These cavity modes have intrinsic loss rate $\kappa_{{\rm i}}$ and
are coupled to each other at rate $h$ (e.g., due to internal scattering)\cite{Spillane03}. 
Input to and output from the internal modes
of the cavity are provided by a tapered fiber with external
coupling rate $\kappa_{{\rm ex}}$, where the fields of the tapered
fiber are denoted by $\{a_{{\rm in}},a_{{\rm out}},b_{{\rm in}},b_{{\rm out}}\}$
in Fig. 1(a). For single-sided excitation as in our current experiment,
the input modes have coherent amplitudes \begin{equation}
\langle a_{{\rm in}}\rangle=-\frac{i\mathcal{E}_{{\rm p}}e^{-i\omega_{p}t}}{\sqrt{2\kappa_{{\rm ex}}}},\quad\langle b_{{\rm in}}\rangle=0.\end{equation}

The Hamiltonian, master equation, and input-output relations for the
system depicted in Fig. 1(a) can be found in Refs.~\cite{Aoki06,Dayan08,Dayan08SI}.
When the total cavity decay rate, $\kappa=\kappa_{{\rm i}}+\kappa_{{\rm ex}}$,
is much larger than any other rate in the system, the cavity modes
can be adiabatically eliminated from the system dynamics~\cite{Gardiner99}.
The system can then be described by effective optical Bloch equations
for a two-level atom with the cavity-enhanced atomic decay rate $\Gamma=\gamma(1+2\mathcal{C})$,
where $\mathcal{C}=\frac{2|g_{{\rm tw}}|^{2}\kappa}{\gamma(\kappa^{2}+h^{2})}$
is the cooperativity parameter for a single atom and $\gamma$ is
the rate of atomic decay to modes other than $a,b$.

In the overcoupled regime of toroid-taper coupling, $\kappa_{{\rm ex}}\gg(\kappa_{{\rm i}},h)$,
the analytic expressions for steady-state transmitted $I_{T}=\langle a_{{\rm out}}^{\dagger}a_{{\rm out}}\rangle_{{\rm ss}}$
and reflected $I_{R}=\langle b_{{\rm out}}^{\dagger}b_{{\rm out}}\rangle_{{\rm ss}}$
fluxes, and for the second-order correlation functions, $g_{{\rm T}}^{(2)}(\tau)=\langle a_{{\rm out}}^{\dagger}a_{{\rm out}}^{\dagger}(\tau)a_{{\rm out}}(\tau)a_{{\rm out}}\rangle_{{\rm ss}}/I_{T}^{2}$
and $g_{{\rm R}}^{(2)}(\tau)=\langle b_{{\rm out}}^{\dagger}b_{{\rm out}}^{\dagger}(\tau)b_{{\rm out}}(\tau)b_{{\rm out}}\rangle_{{\rm ss}}/I_{R}^{2}$,
can be greatly simplified. Specifically, at zero detuning (i.e., $\omega_{{\rm A}}=\omega_{{\rm C}}=\omega_{{\rm p}}$)
and in the weak driving limit, the transmission and reflection coefficients
are given by~\cite{Dayan08SI}
\begin{eqnarray}
T(\Delta=0)\equiv I_{T}(0)/I_{T}(\Delta\gg\kappa)\equiv T_{0} & \simeq & \frac{1}{(2\mathcal{C}+1)^{2}},\\
R(\Delta=0)\equiv I_{R}(0)/I_{T}(\Delta\gg\kappa)\equiv R_{0} & \simeq & \left(\frac{2\mathcal{C}}{2\mathcal{C}+1}\right)^{2},
\end{eqnarray}
 while second-order intensity correlations at zero time delay are
found to be \begin{eqnarray}
g_{{\rm T}}^{(2)}(\tau=0) & \simeq & (4\mathcal{C}^{2}-1)^{2},\label{eq:g2FF}\\
g_{{\rm R}}^{(2)}(\tau=0) & \simeq & 0.\label{eq:g2BB}\end{eqnarray}

The physical interpretation of these results is as follows. In steady
state and with $\mathcal{C}\gg1$, the field radiated by the atomic polarization
$\sigma_{-}^{SS}$ interferes destructively with the intracavity field
from $a_{in}$, resulting in small transmission $T_{0}\ll1$. But
$\sigma_{-}^{SS}$ also coherently drives the intracavity field $b$,
thereby generating a backward propagating field $b_{out}$, leading
to reflection $R_{0}\rightarrow1$~\cite{Shen05}.
On the other hand, the overcoupled cavity without the atomic polarization
(i.e., $\mathcal{C}=0$) has $T_{0}\simeq1$ and $R_{0}\simeq0$. Strong optical
nonlinearity of the atom-toroid system gives rise to a dynamical switching
between these two limits conditioned upon photon detection. (1) $T_{0}\simeq0$,
$R_{0}\simeq1$ due to the field from $\sigma_{-}^{SS}$, and (2)
$T_{0}\simeq1$, $R_{0}\simeq0$ due to the conditional atomic polarization
$\sigma_{-}^{C}$ following photon detection~\cite{Carmichael93}.

This description is substantiated by examination of $g_{T,{\rm R}}^{(2)}$.
Detection of a `first' photon for the reflected field $b_{out}$ projects
the atom to the ground state, which precludes the detection of a `second'
photon for $b_{out}$~\cite{Dayan08,Carmichael93}. Hence the reflected
light exhibits sub-Poissonian photon statistics and photon antibunching.
Similarly, photon detection for the transmitted light $a_{out}$ 
results in atomic projection that depends on $\mathcal{E}_{{\rm p}}$ and $\mathcal{C}$~\cite{Carmichael93}, 
with then $a_{out}$
displaying photon bunching and super-Poissonian statistics. The net
effect is the routing of single photons into the backward direction
$b_{out}$ and remaining excess photons into the forward field $a_{out}$.

Figures~1(b-d) present theoretical results for transmission and reflection
spectra $T(\Delta)$, $R(\Delta)$ and intensity correlations $g_{{\rm T}}^{(2)}(0)$, $g_{{\rm R}}^{(2)}(0)$
in the weak-driving limit for the parameters of our experiment: $(g_{{\rm tw}},\kappa_{{\rm ex}},\kappa_{{\rm i}},h)/2\pi=(50,300,20,10)$
MHz. Although not deeply in the overcoupled regime, efficient transmission
and reflection can be achieved, with $T_{0}\approx0.005$, $R_{0}\approx0.7$ with an intracavity atom, and $T_{0}\approx0.8$, $R_{0}\approx0.003$ without the atom.

A schematic of our experiment is shown in Fig.~1(e) and is similar
to that described in Refs.~\cite{Aoki06,Dayan08}. However, our new apparatus 
has two major improvements that allow us to reproducibly tune $\kappa_{\rm ex}$ 
to the desired overcoupled value and to maintain a small $\kappa_{\rm i}$. 
Firstly, the silicon chip
supporting an array of SiO$_{2}$ toroidal cavities and the
tapered fiber for input-output coupling are mounted on piezoelectric-driven
stages that are placed inside the
main chamber. The combination of compact, rigid design and good passive
vibration-isolation enable us to tune the separation between the taper
and toroid and to achieve stable coupling between the evanescent fields
of the taper and toroid without having the taper contact the surface
of the toroid.

Secondly, in order to avoid degradation of quality factor for the
toroids caused by Cesium contamination, our new apparatus
has a separate {}``MOT chamber'' where Cesium atoms are magneto-optically
cooled and trapped, then loaded into an optical conveyor belt~\cite{Kuhr01}
and transferred into the {}``cQED chamber'' through a differential
pumping tube. The optical conveyor is formed by two counter-propagating
beams red detuned 100 GHz below the D2 line
of Cesium to form a standing-wave dipole-force trap. By independently
shifting the frequencies of each beam with acousto-optic modulators,
atoms loaded from the MOT are conveyed 25 cm from MOT to cQED chamber
and positioned directly above a particular toroid on the silicon chip.

The resonance frequency $\omega_{C}$ of the toroid is tuned 
via the chip temperature near
the frequency $\omega_{A}$ of the $6S_{1/2},F=4$ $\rightarrow$
$6P_{3/2},F^{\prime}=5$ transition of Cesium, with $|\omega_{A}-\omega_{C}|/2\pi\lesssim 30$
MHz. The frequency $\omega_{p}$ of the probe beam $\mathcal{E}_{{\rm p}}$
is locked to coincide with $\omega_{A}$, with $|\omega_{A}-\omega_{p}|/2\pi\lesssim 1$
MHz. The probe is sent into the taper via
the 1\% port of 99/1 fiber beamsplitter (Fig. 1(e)). The 99\%
port of this beamsplitter is connected to a 50/50 beamsplitter followed
by two single-photon counting-modules (SPCMs), $D_{1},D_{2}$, to
detect the reflected field $b_{out}$. Similarly, the transmitted
field $a_{out}$ in the taper is detected with another pair of SPCMs,
$D_{3},D_{4}$. Photoelectric events from the four SPCMs are time
stamped and recorded for analysis.

The physical properties of the new toroids for our current experiment
are similar to those in Ref.~\cite{Dayan08}. The major diameter
$D\approx25 \mu$m and minor diameter $d\approx6 \mu$m, leading to an
effective atom-cavity coupling rate of $g_{{\rm eff}}/2\pi\approx70$
MHz ($g_{{\rm tw}}/2\pi\approx50$ MHz) \cite{Aoki06,g}. Combined with the measured values for
$(\kappa_{{\rm ex}},\kappa_{{\rm i}},h)$ and with $\gamma/2\pi=5.2$
MHz, we calculate the single-atom cooperativity parameter $\mathcal{C}=3.0$
and the cavity-enhanced atomic decay rate $\Gamma/2\pi=36.4$ MHz.
Note that for large toroid-taper separation with $\kappa_{{\rm ex}}\simeq\kappa_{i}$, 
our system is in the strong coupling regime of cQED with $g_{\rm eff}>\kappa,\gamma$.

About 10$^{7}$ atoms with temperature $T\approx100 \mu$K are dropped
from a height of $0.5$ mm above the toroid by turning off the dipole
trap beams. As shown in Fig.~1(c), the reflected intensity $I_{R}$
associated with $b_{out}$ should be near zero when there is no atom
in the cavity. A falling atom that interacts with the evanescent fields
of the cavity generates an increase in $I_{R}$ for the duration of
the atomic transit and enables single transit events to be observed
with high signal-to-noise ratio \cite{transit}. While transiting atoms are sufficient 
for these experiments due to our insensitivity to atomic position, 
in the future we recognize the need to trap and localize single atoms 
near the toroid to allow continuous operation of the system 
\cite{McKeever04,Keller04,Hijelkema07,Boozer07}.

\begin{figure}[t,b]
\includegraphics[width=8.6cm]{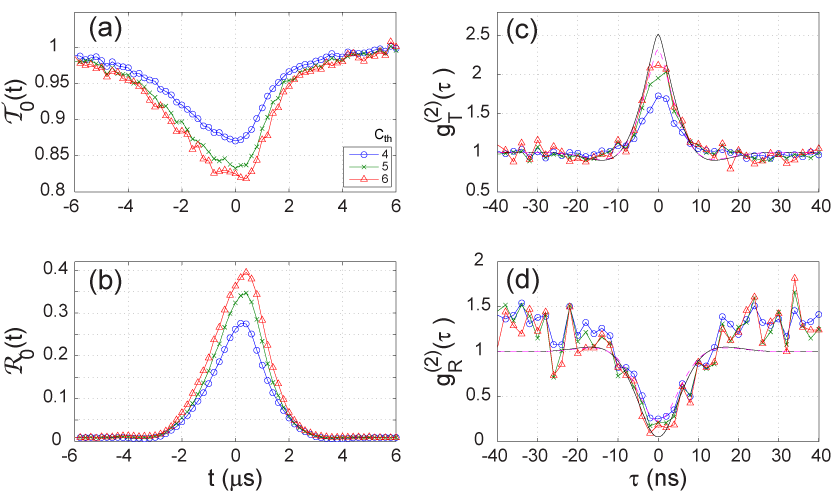}
\caption{\label{fig:2} (a,b) Average atom transit signals for (a) transmission
$\mathcal{T}_{{\rm 0}}(t)$ and (b) reflection $\mathcal{R}_{0}(t)$
of the probe field. As shown in the inset in (a), the transit selection
criteria are set to be $C_{{\rm th}}=4,5,6$, where in all cases,
$\Delta t_{{\rm atom}}=4 \mu$s. (c,d) The intensity correlation functions
$g_{{\rm T,R}}^{(2)}(\tau)$ for the transmitted field $a_{out}$
and the reflected field $b_{out}$. For (a-d), $\bar{n}=0.093$ photons.
Solid lines are a theoretical calculation using the
parameters $(g_{\rm tw}, \kappa_{\rm ex}, \kappa_{\rm i}, h)/2\pi= (50,300,20,10)$ MHz. 
Dashed lines are the same calculation with 4\% background counts.
}
\end{figure}

We first fix the probe power $|\mathcal{E}_{{\rm p}}|^{2}$ such that
the intracavity photon number $\bar{n}=0.093$ for the empty cavity.
Atom transit events are extracted from the records of photoelectric
counts $C_{1,2}(t)$ from $b_{out}$ at detectors $D_{1},D_{2}$ by
the following procedure. For an atom transit event, we require the
sum of the counts over a time $\Delta t_{{\rm atom}}$, $S_{j}\equiv\sum_{t_{i}=t_{j}}^{t_{j}+\Delta t_{{\rm atom}}}[C_{1}(t_{i})+C_{2}(t_{i})]$,
to be equal or greater than a threshold count $C_{{\rm th}}$. After
applying this selection criterion to $C_{1,2}(t)$, we determine the
time origin $t=0$ for each transit event by the temporal mean of the 
counts in $\Delta t_{{\rm atom}}$ and extract counts in a time window
of $\pm6\mu$s around $t=0$ for further analysis.

Figure~2(a,b) show the transmitted $\mathcal{T}_{{\rm 0}}(t)$ and
reflected $\mathcal{R}_{0}(t)$ signals averaged over transit events
with transit selection criteria of $C_{{\rm th}}=4,5,6$ and $\Delta t_{{\rm atom}}=4 \mu$s,
where $\mathcal{T}_{{\rm 0}}(t)$ and $\mathcal{R}_{0}(t)$ are normalized
to the resonant transmission of the empty cavity, $T_0 \simeq 0.83$. Atom-cavity coupling
results in reduced transmission and increased reflection~\cite{asymmetry}. 
The second-order correlation functions $g_{{\rm T,R}}^{(2)}(\tau)$ for the transmitted
and reflected fields $a_{out},b_{out}$ are calculated in the same
manner as in Refs.~\cite{Dayan08,Dayan08SI} and plotted in Fig.~2(c,d). Photon
bunching and antibunching are clearly observed for $a_{out},b_{out}$
in the forward and backward directions, respectively. The recovery
time $\tau_{R}$ to steady state is set by the cavity-enhanced atomic decay
rate $\Gamma=\gamma(1+2\mathcal{C})$, with $\tau_{R}\sim1/\Gamma$ \cite{Dayan08,Dayan08SI}.

These observations are consistent with the theoretical predictions
in Fig.~1(b-d) for $\Delta=0$, thereby demonstrating a functional
photon router. Note that the traces with larger threshold $C_{{\rm th}}$
show more significant effects of atom-cavity coupling. This is because
we have fewer {}``falsely detected transits'' (i.e., time windows
that have no atom but nevertheless satisfy the selection criterion
$S_{{\rm j}}\geq C_{{\rm th}}$ because of background light) and reduced
contributions from transits with smaller $g$.

\begin{figure}[t,b]
\includegraphics[width=8.6cm]{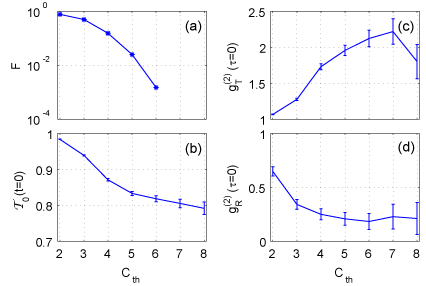}
\caption{\label{fig:3} (a) False detection ratio $F$, (b) transmitted signal
$\mathcal{T}_{{\rm 0}}(t=0)$ at the center of an atomic transit,
and (c,d) intensity correlation functions $g_{{\rm T,R}}^{(2)}(\tau~=~0)$
at zero time delay for the transmitted $T$ and reflected $R$ light
as functions of the threshold $C_{{\rm th}}$ for the selection of
atom transits. In all cases, $\Delta t_{{\rm atom}}=4 \mu$s and $\bar{n}=0.093$.}
\end{figure}

To explore this point further, in Fig.~3(a) we plot the ratio $F$
of false to total atomic transit events, $F\equiv \mathcal{R}_{0}^{no}(t=0)$/$\mathcal{R}_{0}(t=0)$,
where $\mathcal{R}_{0}^{no}$ is the measured transit signal for the reflected
field as in Fig. 2(b), but now with no atoms present. Clearly, $F$
decreases as $C_{{\rm th}}$ increases. Correspondingly, in (b) - (d), $\mathcal{T}_{0}$, $g_{{\rm T}}^{(2)}$,
and $g_{{\rm R}}^{(2)}$ show more significant effects of
atom-cavity coupling with increasing $C_{{\rm th}}$, albeit with
larger error bars due to the diminishing number of transit events.
The value $C_{{\rm th}}=5$ ($F\simeq0.025$)
provides a reasonable compromise between decreasing signal-to-noise
ratio and reduced contamination from false transit events.

Finally, in order to investigate the saturation behavior of our photon
router, we vary the intracavity photon number by way of the probe intensity
$|\mathcal{E}_{{\rm p}}|^{2}$. Figure~4 displays the
measured values of $\mathcal{T}_{{\rm 0}}(t=0)$
and $g_{{\rm R}}^{(2)}(\tau=0)$ as functions of $\bar{n}$. For each
value of $\bar{n}$, we set the selection threshold $C_{{\rm th}}$
so that the false detection ratio $F<0.05$. The point for the lowest value of $\bar{n}$ for $g_{{\rm R}}^{(2)}(\tau=0)$
is omitted from (b) due to its poor signal-to-noise ratio.
In (a), the transmission $\mathcal{T}_{{\rm 0}}(t=0)$ shows clear
saturation of the atom-cavity coupling, from $\mathcal{T}_{{\rm 0}}\approx0.2$
at $\bar{n}\approx0.01$ to $\mathcal{T}_{{\rm 0}}\approx0.8$ at
$\bar{n}\approx0.7$. Importantly, in (b) $g_{R}^{(2)}(\tau=0)<1$ over a
wide range of intracavity photon number, $0.01\lesssim\bar{n}\lesssim0.7$.

From the measured $\mathcal{T}_0$
and known parameters $(g_{\rm tw},\kappa_{{\rm ex}},\kappa_{{\rm i}},h)$,
we estimate the efficiency $\xi$ for single-photon throughput 
and obtain $\xi \approx 0.6$ for the lowest $\bar{n} \approx 0.012$, with $\xi\rightarrow0.7$ for $\bar{n}\rightarrow0$.
This is consistent with the value $\xi \approx 0.5 \pm 0.2$ directly inferred 
from the distribution of $\mathcal{R}_0$ for the individual transits. As evidenced in Fig.
2(a, c), multiple photons are efficiently transmitted into the output
channel $a_{out}$.

\begin{figure}[t,b]
\includegraphics[width=8.6cm]{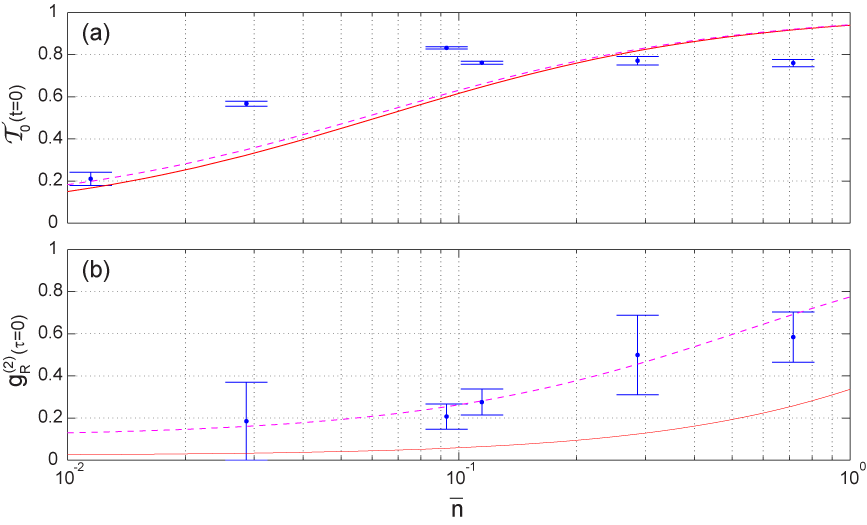}
\caption{\label{fig:4} (a) Transmitted signal $\mathcal{T}_{{\rm 0}}(t=0)$
at the center of an atomic transit and (b) intensity correlation
function $g_{{\rm R}}^{(2)}(\tau=0)$ at zero time delay for the reflected light for various values of intracavity
photon number $\bar{n}$. Points are experimental data averaged over
individual transit events. Solid lines are from
a theoretical calculation with the parameters
$(g_{\rm tw}^{min}, g_{\rm tw}^{max}, \kappa_{\rm ex}, \kappa_{\rm i}, h)/2\pi=(35,65,300,20,10)$ MHz where
instead of a single value of $g_{\rm tw}$ we use an average over $g_{\rm tw}^{min}$ to 
$g_{\rm tw}^{max}$.  Dashed lines are the same calculation, but with the assumption of
background counts of 4\% of the signal.}
\end{figure}

The solid lines in Fig. 4 are from a simplified theoretical model
based on adiabatic elimination of the cavity \cite{Dayan08SI}.
The curves are obtained from an average over the atomic azimuthal
position $kx$ around the major circumference of the toroid.
Curves for particular values of $kx$ would be indistinguishable on
the scale of the plot. This insensitivity to $kx$ is due to the small 
coupling between the counterpropagating modes $a$ and $b$, 
characterized by the ratio $h/\kappa\simeq0.03\ll1$, that limits 
the formation of a standing wave~\cite{Dayan08SI}.

To account for background light, the dashed lines in Fig.~4 include
background counts at a level of 4\% of the signal.  This leads to good
agreement for the background-sensitive parameter $g_{{\rm R}}^{(2)}(\tau=0)$ in Fig.~4(b).
There still remains a discrepancy for $\mathcal{T}_0$ in Fig. 4(a) given the statistical
error bars, even when we account for variation in $g_{\rm tw}$ by averaging over a
range of values around $g_{\rm tw} = 50$ MHz ( $g_{\rm tw} = 35$ MHz to  $g_{\rm tw} = 65$ MHz).  
Thus, we must attribute the discrepancy in (a) to a systematic uncertainty in the
experiment.

In summary, we have realized a photon router using single Cesium atoms
coupled to a microtoroidal cavity in the overcoupled regime. In comparison
to our previous photon turnstile~\cite{Dayan08}, this photon router
is robust against azimuthal position of the atom ($kx$) and taper-toroid
distance ($\kappa_{{\rm ex}}$), in addition to the atom-cavity coupling  
strength ($g(\overrightarrow{r})$). The current system already has
throughput efficiency for single photons of $\xi\approx 0.6$\ and
it is projected to reach $\xi> 0.999$\ through use of smaller toroid 
diameter~\cite{Spillane05} and improvement of the
intrinsic quality factor of the cavity to $Q\approx10^{10}$~\cite{Vernooy98}.
The realization of strong interactions of single photons and atoms
together with efficient input-output provides an enabling capability
for the realization of quantum networks and investigations of quantum
many-body systems built component by component.

We gratefully acknowledge the contributions of S. Kelber and E. Wilcut-Connolly. 
This work is supported
by the NSF PHY-0652914, by IARPA via ARO, and by NGST. ASP thanks the Marsden Fund 
of the Royal Society of New Zealand for support.

{\scriptsize Permanent addresses: $^{a}$ TA - Department of Physics, Kyoto University, Kyoto, Japan, and 
PRESTO, Japan Science and Technology Agency, Saitama, Japan; $^{b}$ ASP - Department of Physics, University of Auckland, Auckland, New
Zealand; $^{c}$ BD - Department of Chemical Physics, Weizmann Institute of Science, Rehovot, Israel.}{\scriptsize \par}


\begin{thebibliography}{99}
\bibitem{g}
The rate of coherent interaction between the atom and the modes $a$,
$b$ is given by $g_{{\rm tw}}=g_{0}^{{\rm tw}}f(\rho,z)e^{\pm ikx}$,
where $\rho$ is the atom-toroid radial distance, $x$ is the atom's
position around the toroid circumference, and $z$ is the vertical
coordinate.
\bibitem{transit}
For our previous work~\cite{Aoki06,Dayan08}, the taper and toroid were critically coupled with the taper
in contact mode. Atom transits were detected as increases in the
transmitted intensity $I_{T}$ for the field $a_{out}$.
\bibitem{asymmetry}
The asymmetry evident in Fig.~2(a,b) is a robust feature of our observations. 
We have no definitive explanation, but are investigating various possible causes, 
including surface forces.
\end{thebibliography}
\end{document}